\documentclass[twocolumn,english,nobibnotes,nofootinbib,superscriptaddress]{revtex4-1}
\usepackage[T1]{fontenc}
\usepackage[a4paper]{geometry}
\usepackage{float}
\usepackage{color}
\usepackage{xfrac}
\usepackage{textcomp}
\usepackage{amsmath}
\usepackage{amssymb}
\usepackage{graphicx}
\usepackage{esint}
\usepackage{natbib}
\usepackage[normalem]{ulem}
\makeatletter
\@ifundefined{textcolor}{}
{%
 \definecolor{BLACK}{gray}{0}
 \definecolor{WHITE}{gray}{1}
 \definecolor{RED}{rgb}{1,0,0}
 \definecolor{GREEN}{rgb}{0,1,0}
 \definecolor{BLUE}{rgb}{0,0,1}
 \definecolor{CYAN}{cmyk}{1,0,0,0}
 \definecolor{MAGENTA}{cmyk}{0,1,0,0}
 \definecolor{YELLOW}{cmyk}{0,0,1,0}
}

\begin{document}

\author{A. Mre\'{n}ca-Kolasi\'{n}ska}
\affiliation{AGH University of Science and Technology, Faculty of Physics and
Applied Computer Science,\\
 al. Mickiewicza 30, 30-059 Kraków, Poland}
\affiliation{NEST, Istituto Nanoscienze-CNR and Scuola Normale Superiore, \\
Piazza San Silvestro 12,
56127 Pisa, Italy}

\author{S. Heun}
\affiliation{NEST, Istituto Nanoscienze-CNR and Scuola Normale Superiore, \\
Piazza San Silvestro 12,
56127 Pisa, Italy}

\author{B. Szafran}
\affiliation{AGH University of Science and Technology, Faculty of Physics and
Applied Computer Science,\\
 al. Mickiewicza 30, 30-059 Kraków, Poland}

\title{Aharonov-Bohm interferometer based on n-p junction in graphene nanoribbon}
\begin{abstract}
We demonstrate that the phenomenon of current confinement along graphene n-p junctions at high magnetic fields
can be used to form an Aharonov-Bohm interferometer.
The interference system exploits a closed n-p junction that can be induced by a floating gate within the sample, and coupling of the junction currents with the edge currents in the quantum Hall regime. Operation of the device requires current splitting at the edge and the n-p junction contacts which is found for armchair ribbons at low Fermi energy.
\end{abstract}
\maketitle

\section{Introduction}

One of the fascinating properties of graphene is that one can induce regions of hole or electron conductivity with potentials applied to external gates, without the need for chemical doping~\cite{Neto}.
The electrostatic control of the position of the chemical potential with respect to the Dirac point allows for a precise definition of n-p junctions within the sample.
A well known feature of such junctions is Klein tunnelling \cite{Katsnelson06}: in the absence of an external magnetic field, the n-p junction is transparent for electrons under normal incidence.
The angular dependence of the phenomenon allows for the observation of Fabry-P\'erot interference in n-p-n junctions \cite{Shytov08,FP,FP2}.
In high magnetic fields, n-p junctions \cite{ob,wy} form waveguides for electrons \cite{Ghosh,Cresti,Rickhaus}, similar to the edge currents flowing in a two-dimensional electron gas in the quantum Hall regime.
In a semi-classical picture, the carriers move along the junction on snake orbits \cite{Williams,Mueller,Carmier10,Oroszlany08,Zarenia13,skippingsnakes,Rickhaussnake}.
Current oscillations along a n-p interface due to snake orbits have been very recently observed in gated graphene \cite{Rickhaussnake,Taya}.
Also the conductance of graphene n-p junctions in the quantum Hall regime was studied in a number of papers \cite{Rickhaussnake,Chen,Patel,Milan,ob,wy,Cresti,Tworzydlo,Long}.

In this work we propose an Aharonov-Bohm \cite{AB} quantum Hall interferometer \cite{Halperin11} exploiting the current confinement along a circular n-p junction induced by the tip of an atomic force microscope \cite{sgmr1,Cabosart} placed above an armchair graphene nanoribbon \cite{Waka}. 
The operation of the device is based on the coupling of the edge and junction currents in the quantum Hall regime.
The system is a type of quantum ring. The conductance of open graphene quantum rings \cite{Peetnano,Schelter,Xu,Abelger,Faria,Wurm}
as well as the energy spectra of closed rings  were already evaluated \cite{Costa2014,Hewe2008,Downing11,Pecher,Yannou,Zarenia}. The systems studied by theory
\cite{Peetnano,Schelter,Xu,Abelger,Faria,Wurm,Costa2014,Hewe2008,Downing11,Pecher,Yannou,Zarenia, Rakyta} and experiment \cite{Magdalena,Cabosart,Russo,smirnov} were based on graphene samples with material removed from the center of the ring.
The interplay of the Klein and Aharonov-Bohm effects was studied in etched nanorings \cite{Schelter}.  Conductance of quantum rings in Corbino geometry was also discussed \cite{Rycerz,Rycerz2}.

By exploiting the current confinement in graphene n-p junctions, we gain the possibility to obtain a fully controllable device, in contrast to etched ring-shaped devices with fixed geometry \cite{Ismail,Timp,Strambini}.  We take advantage of the unique nature of graphene to attain a n-p junction with adjustable size and position, as opposed to the n-p junctions in III-V semiconductor structures which have a geometry predefined by the doping profile.

The conductance evaluated in the present work for a circular n-p junction induced electrostatically in armchair nanoribbons exhibits a clear Aharonov-Bohm (AB) periodicity in the low Fermi energy regime. The periodicity appears only in the quantum Hall regime. In contrast to the graphene quantum rings studied so far, the power spectrum of which exhibited a number of higher harmonics also in the single mode regime \cite{Wurm}, here we demonstrate that the conductance of systems based on metallic armchair nanoribbons at low Fermi energy possesses the fundamental AB period only. Higher harmonics are present for semiconducting ribbons, for which the contact of the n-p junction to the edge channels acts as a beam splitter.

\section{Theory}

\begin{figure}[htbp]
\includegraphics[width=0.4\paperwidth]{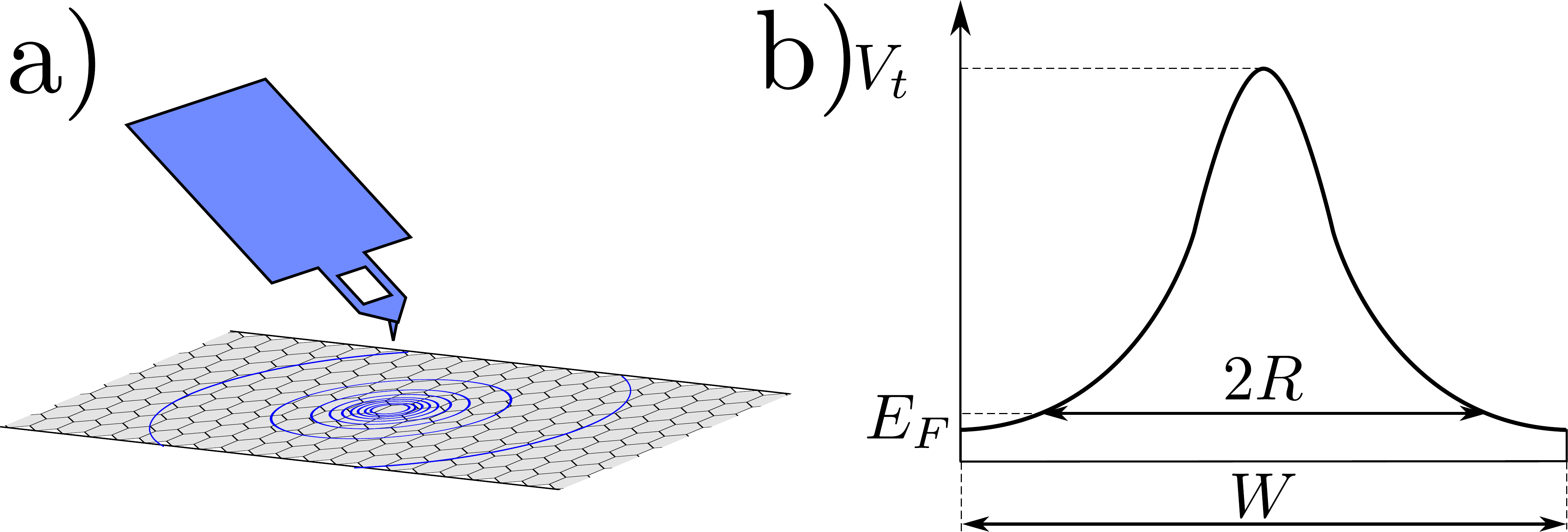}
\caption{(a) Schematic drawing of the studied system: a graphene nanoribbon with
a potential perturbation induced by a floating gate. (b) Tip potential modelled by a Lorentz function Eq.~(\ref{lf}), % {\bf AM: czym jest W}
and the effective diameter $2R$ of the n-p junction for a given Fermi energy $E_F$. $W$ is the width of the ribbon.
}\label{schemat}
\end{figure}

We consider an armchair nanoribbon with the tip floating above [Fig.~\ref{schemat}(a)] and use a tight binding Hamiltonian for $\pi$ electrons
\begin{eqnarray}
H=\sum_{\{i,j\}}(t_{ij} c_i^\dagger c_j+h.c.)+\sum_i V({\bf r}_i) c_i^\dagger c_i, \label{dh}
\end{eqnarray}
with the nearest neighbour hopping parameters including the Peierls phase, $t_{ij}=t\exp\left[\frac{2\pi i}{\Phi_0}\int_{{\bf r}_i}^{{\bf r}_j}{\bf A}\cdot {\bf dl}\right]$,
where $t=-2.7$ eV, $\Phi_0={h}/{e}$ is the flux quantum, and ${\bf r}_i$ is the position of the i-th atom.
We assume that the external magnetic field is applied perpendicular to the plane of confinement ${\bf B}=(0,0,B)$ and use the Landau gauge ${\bf A}=(-By,0,0)$.
The present calculations cover a range of magnetic fields from $B=0$ to the quantum Hall regime.
$V({\bf r})$ in Eq.~(\ref{dh}) stands for the potential due to the tip.
The Coulomb potential of the charge at the tip is screened by deformation of the two-dimensional electron gas.
The form of the resulting effective potential as calculated by the Sch\"odinger-Poisson modelling \cite{kolasinskiDFT2013} is close to
a Lorentz function
\begin{equation}
V(x,y)=\frac{V_t}{1+\left( (x-x_t)^2+(y-y_t)^2\right)/d^2}, \label{lf}
\end{equation}
where $x_t,y_t$ indicate the tip position, $d$ the width of the effective tip potential, and $V_t$ gives the maximal value of the potential [Fig.~\ref{schemat}(b)].
The latter is determined by the potential applied to the tip, and $d$ depends on the tip-electron gas distance \cite{kolasinskiDFT2013}.

In order to evaluate the conductance we use the Landauer approach and solve the scattering problem for the subbands at the Fermi level.
We consider homogeneous armchair with their axis along $y=0$. For evaluation of the dispersion relation far from the tip scatterer,
we  assume electron eigenstates in the nanoribbon in the Bloch form,
\begin{equation}
\psi^{k_m}_{u,v}=\chi^{k_m}_{v} e^{ik_m u\Delta x},\label{blw}
\end{equation}
where $k_m$ is the wave vector for the $m$-th subband, $\chi_{u,v}^{k_m}$ is a periodic function with the crystal periodicity of the ribbon at the $v$th site in the $u$th elementary cell, and $\Delta x$ is $3a_{cc}$ for armchair nanoribbon, with $a_{cc}=1.42$ \AA.
In presence of the tip the wave functions in the input lead are superpositions of the Bloch functions: the incident one $\psi^{k_{in}}_{in}$, and the
ones backscattered by the tip,
\begin{equation}
 \Psi_{in}^{u,v} =  \sum\limits_l  c_{in}^{l} \psi^{k^+_{in}}_{u,v}+ \sum\limits_l  d_{in}^{l} \psi^{k^-_l}_{u,v},
\label{fun_in}
\end{equation}
where the sum runs over the subbands $l$ with the backscattered current flux flowing from the tip to the left lead.
At the right-hand side of the tip the scattering wave has the form
\begin{equation}
 \Psi_{out}^{u,v} =  \sum\limits_l  c_{out}^{l} \psi^{k^+_l}_{u,v},
\label{fun_out}
\end{equation}
where the summation runs over the subbands carrying the current to the right.
The incident amplitude $c_{in}^{l}$ is set to 1 for each subsequent subband. The backscattered $d_{in}$ and transferred amplitudes $c_{out}$
are evaluated with the quantum transmitting boundary method \cite{Zwierz,Lent,Leng,Kolasinski2014,amr,supplemental}.
The scattering amplitudes with the current fluxes allow
 to evaluate the transfer probability $T_{l}$ for the incident subband as $T_{l}= \sum\limits_m T_{ml}$, where $T_{ml}$
may also include transfer between propagating modes in different valleys.
The 0K conductance is then evaluated as $T=2{G_0}\sum_{l} T_{l}$, with $G_0={e^2}/{h}$. The factor of 2 accounts for the spin degeneracy.

The current flow between the atoms $m$ and $n$, as derived from the Schr\"odinger equation \cite{Wakabayashi}, is

\begin{equation}
 J_{mn} =  \frac{i}{\hbar} \left[ t_{mn} \Psi^*_m \Psi_n - t_{nm} \Psi^*_n \Psi_m \right],
\label{current}
\end{equation}
where $ \Psi_n $ is the wave function at the $n$th atom.
%\textcolor{red}{ \sout{ The probability current flux
%can be evaluated as:}
%\begin{equation}
%\sout{ \phi = \sum\limits_m \sum\limits_{n_m}  J_{mn_m} ,}
%\end{equation}
%\sout{where the first sum runs over the atoms at the end of the channel, and the second sum over their neighbors $n_m$ within the computational box.} }

\begin{figure}[htbp]
\includegraphics[width=0.24\paperwidth]{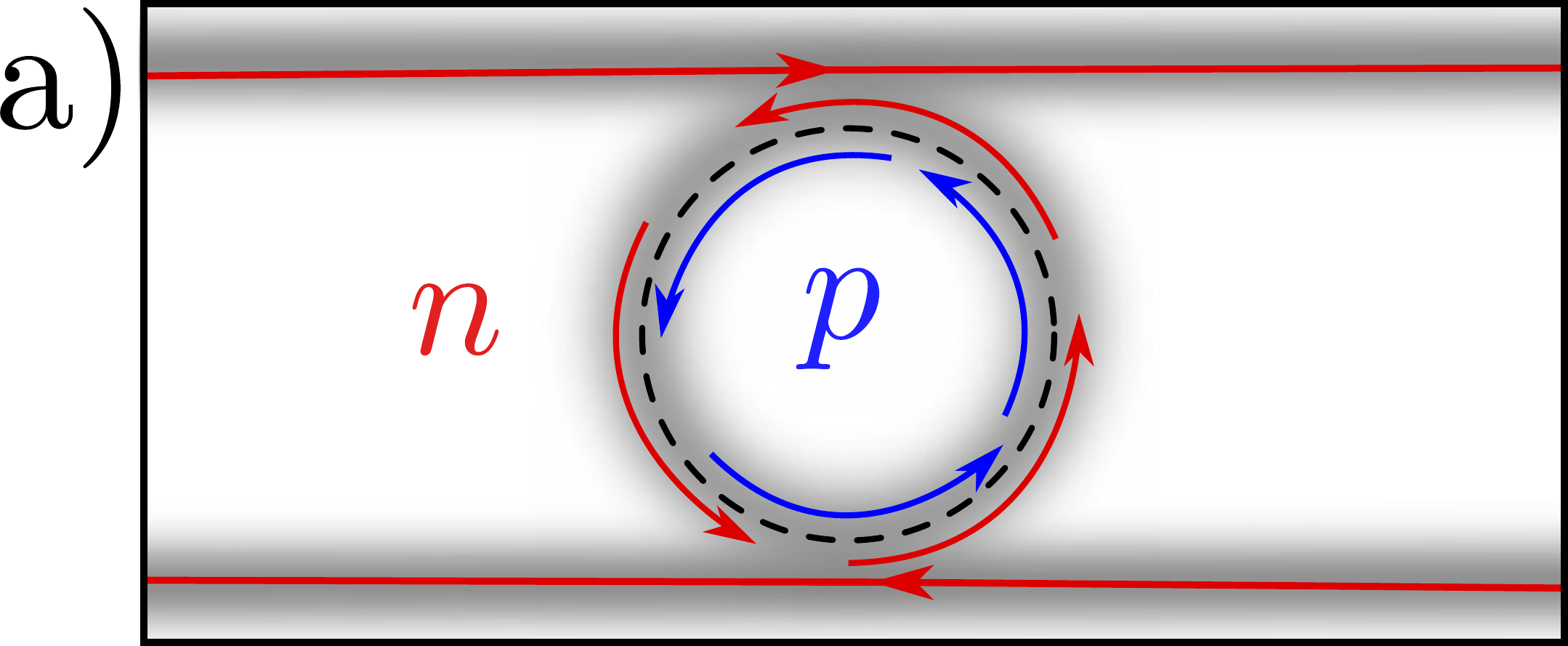}
\includegraphics[width=0.24\paperwidth]{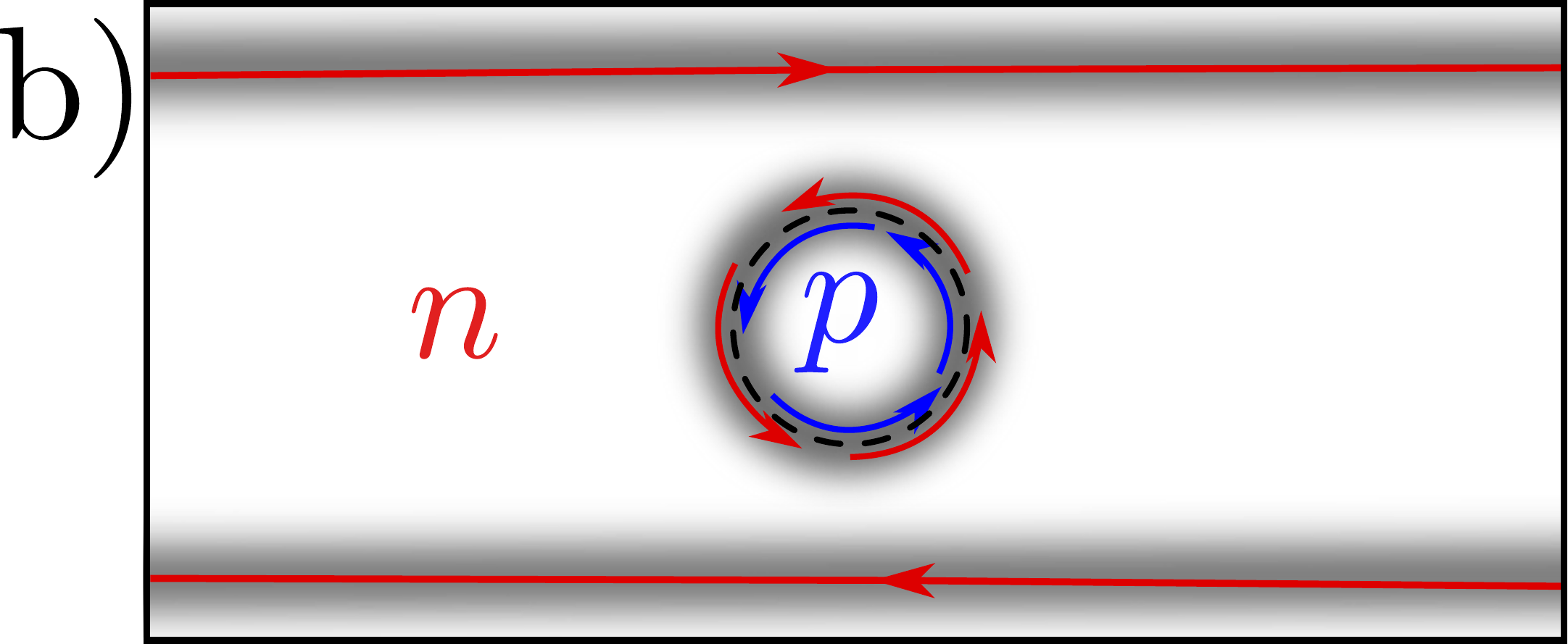}
\caption{ Schematic drawing of the currents in the system in the quantum Hall regime.
(a) Low $E_F$ and edge current coupled to the n-p junction.
 (b) For high $E_F$ the radius of the n-p junction decreases, and the junction is too far from the edge for the edge current to couple to the n-p junction.  %(c) At high magnetic field the
}\label{schemeCurr}
\end{figure}

Figure \ref{schemeCurr} sketches the  current distribution in the quantum Hall regime.
The Fermi energy is set within the conduction band of the ribbon.  The potential of the tip
placed above the center of the ribbon raises the valence band top above the Fermi energy, inducing
a circular region of p-type conductivity.
The current flows near the edge of the ribbon and along the n-p junction. For radius
of the n-p junction large enough to approach the edges of the sample,
 the edge current couples to the n-p junction and flows around the circular p-region [Fig.~\ref{schemeCurr}(a)].
For a higher Fermi energy, the radius of the n-p junction gets smaller, and the coupling of the edge current to the junction becomes weaker.
At some point [Fig.~\ref{schemeCurr}(b)] the edge current can no longer couple to the n-p junction.

\section{Results and discussion}

%\subsection{Armchair ribbons}

\subsection{Low-energy range}

\begin{figure*}[htbp]
\includegraphics[width=0.30\paperwidth]{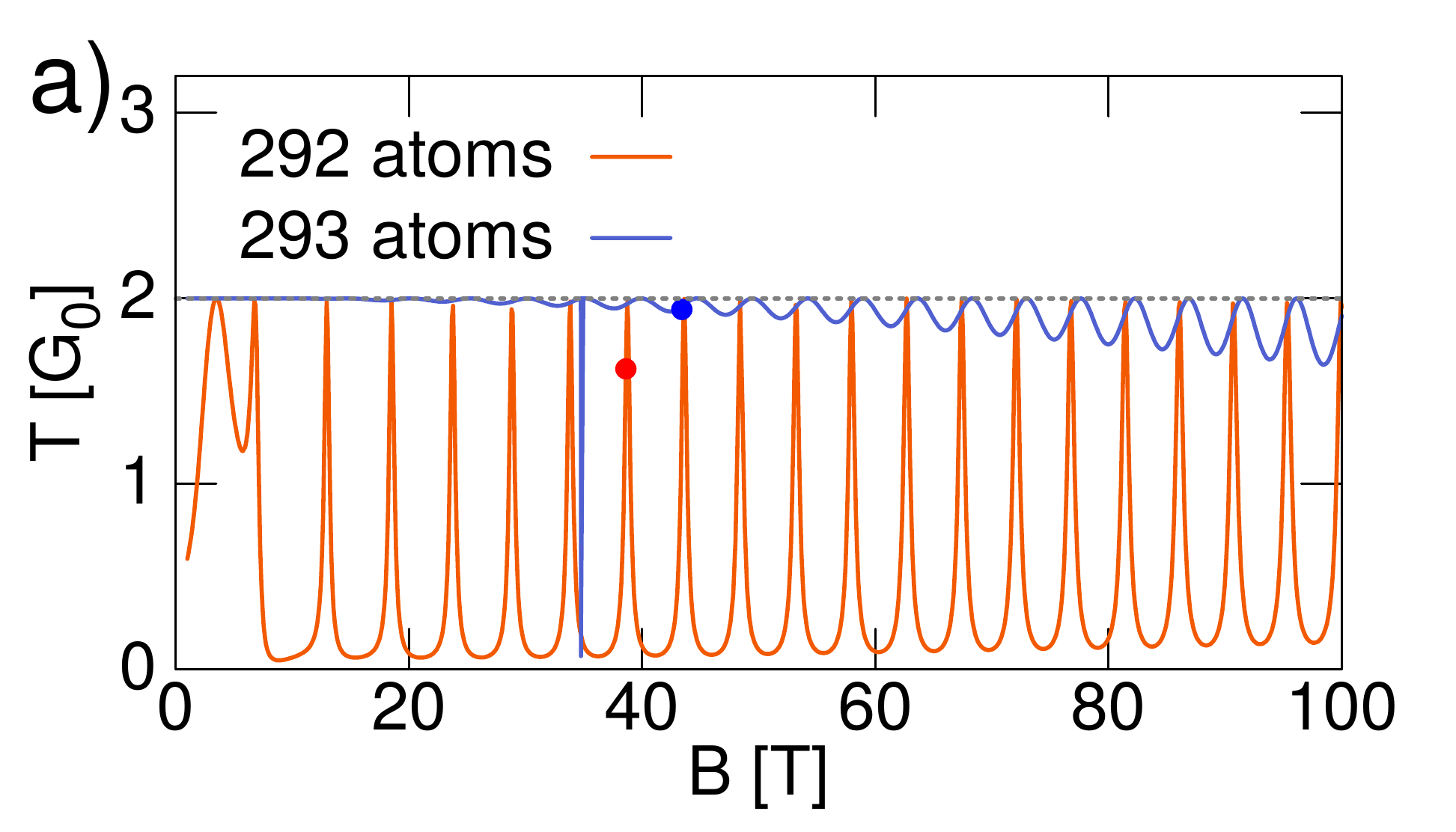}
\includegraphics[width=0.24\paperwidth]{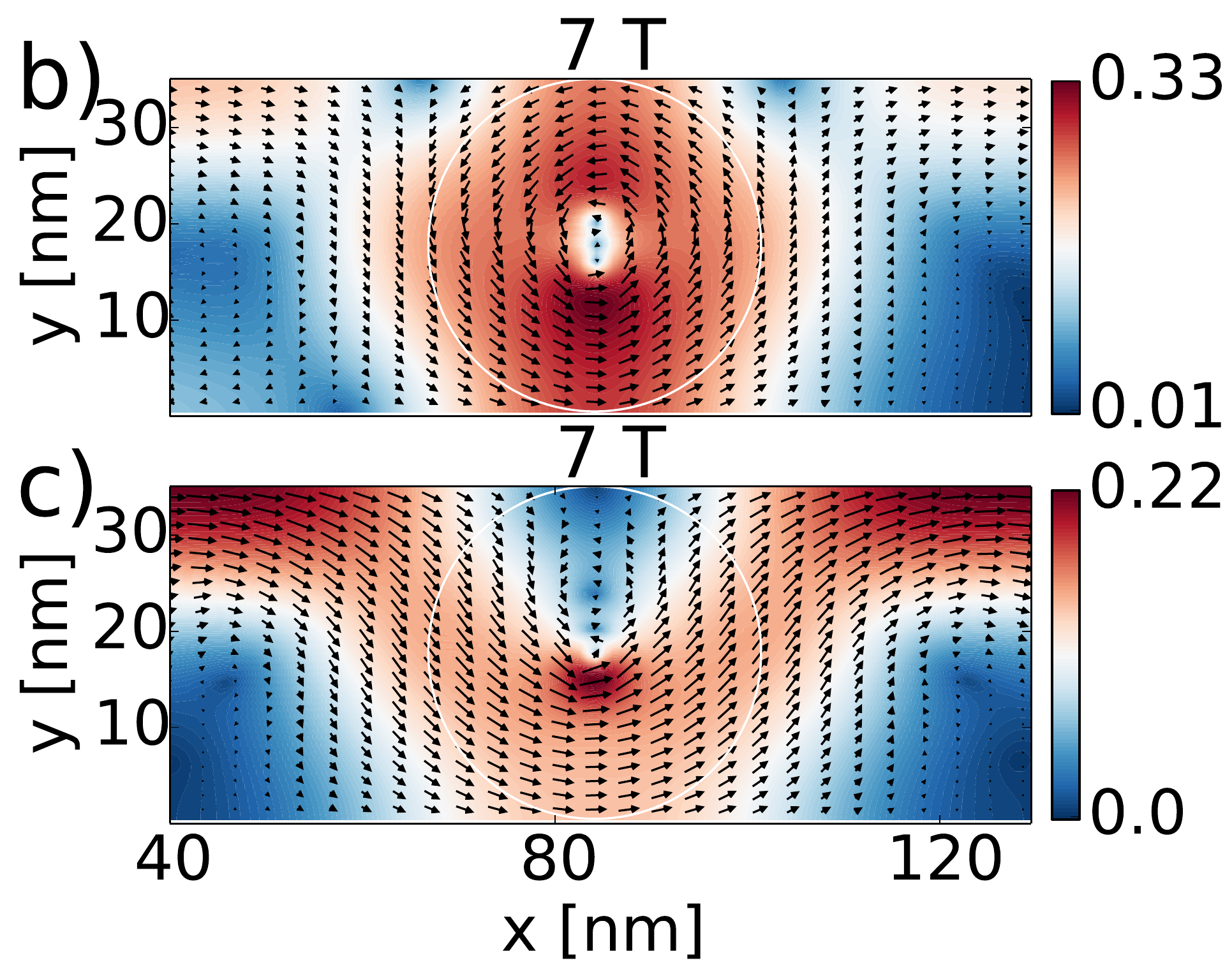}
\includegraphics[width=0.26\paperwidth]{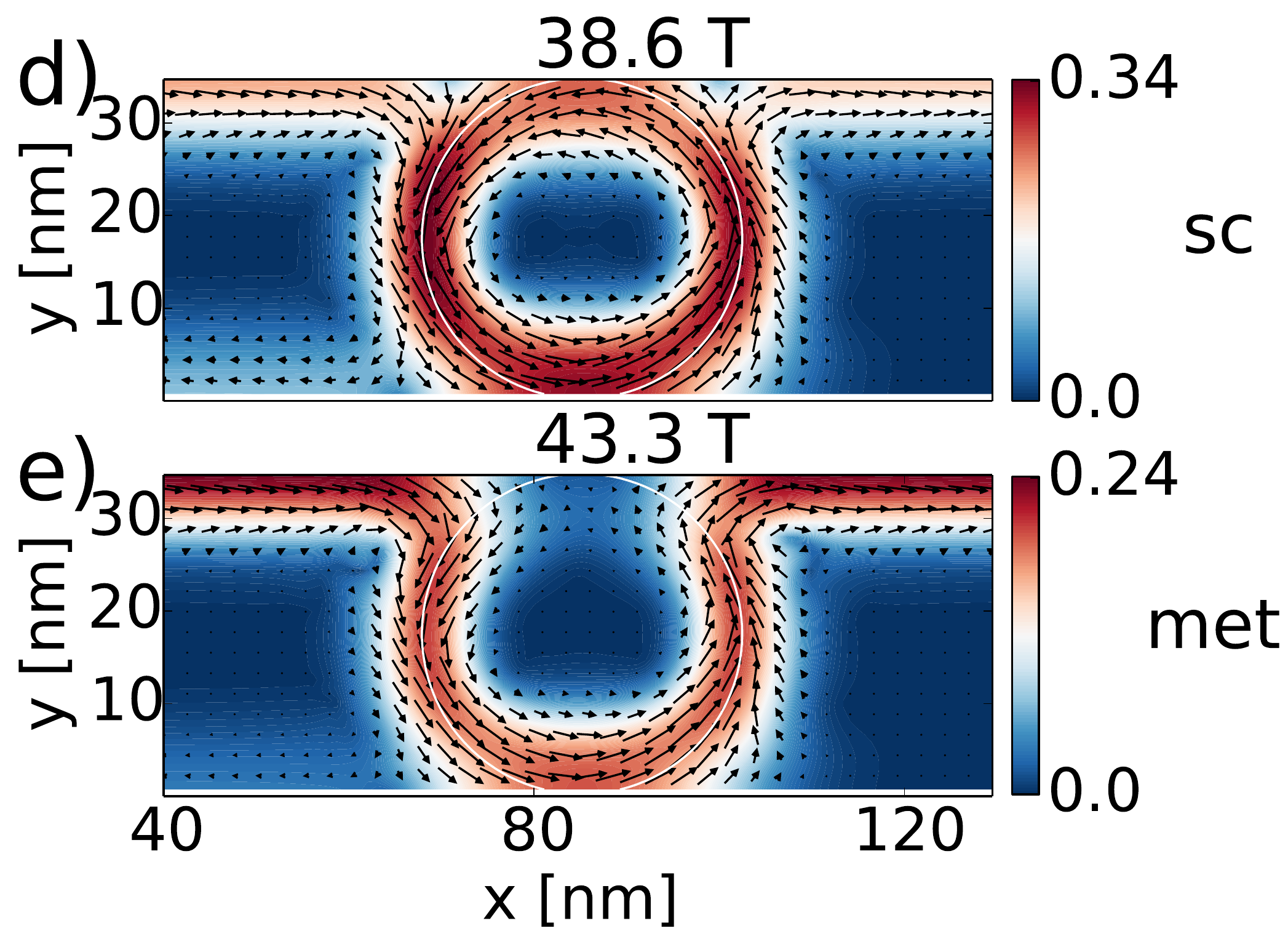}
\caption{
(a) Summed electron transfer probability for armchair nanoribbons (semiconducting with 292 atoms across the channel (35.79 nm) -- orange lines,
and metallic with 293 atoms across the channel (35.92 nm) -- blue lines) in the lowest subband transport conditions for $E_F=30$ meV. The tip is located above the axis of the channel. The applied tip potential is $V_t=400$ meV,
and $d=4.92$ nm. In (b-e) maps of the square root of the current amplitude [current amplitude calculated using Eq.~(\ref{current})] are plotted with orientation of the vector current distribution.
Plots (b,d) were calculated for the semiconducting, and (c,e) for the metallic ribbon.
The external magnetic field is 7T in (b,c) for $B$ below formation of a periodic AB oscillation.
Plot (d) was made for the semiconducting ribbon -- see the orange dot in (a).
Plot (e) corresponds to the metallic ribbon and was taken for the magnetic field marked by the blue dot in (a).
}\label{pwB0304}
\end{figure*}

\begin{figure*}[htbp]
\includegraphics[width=0.30\paperwidth]{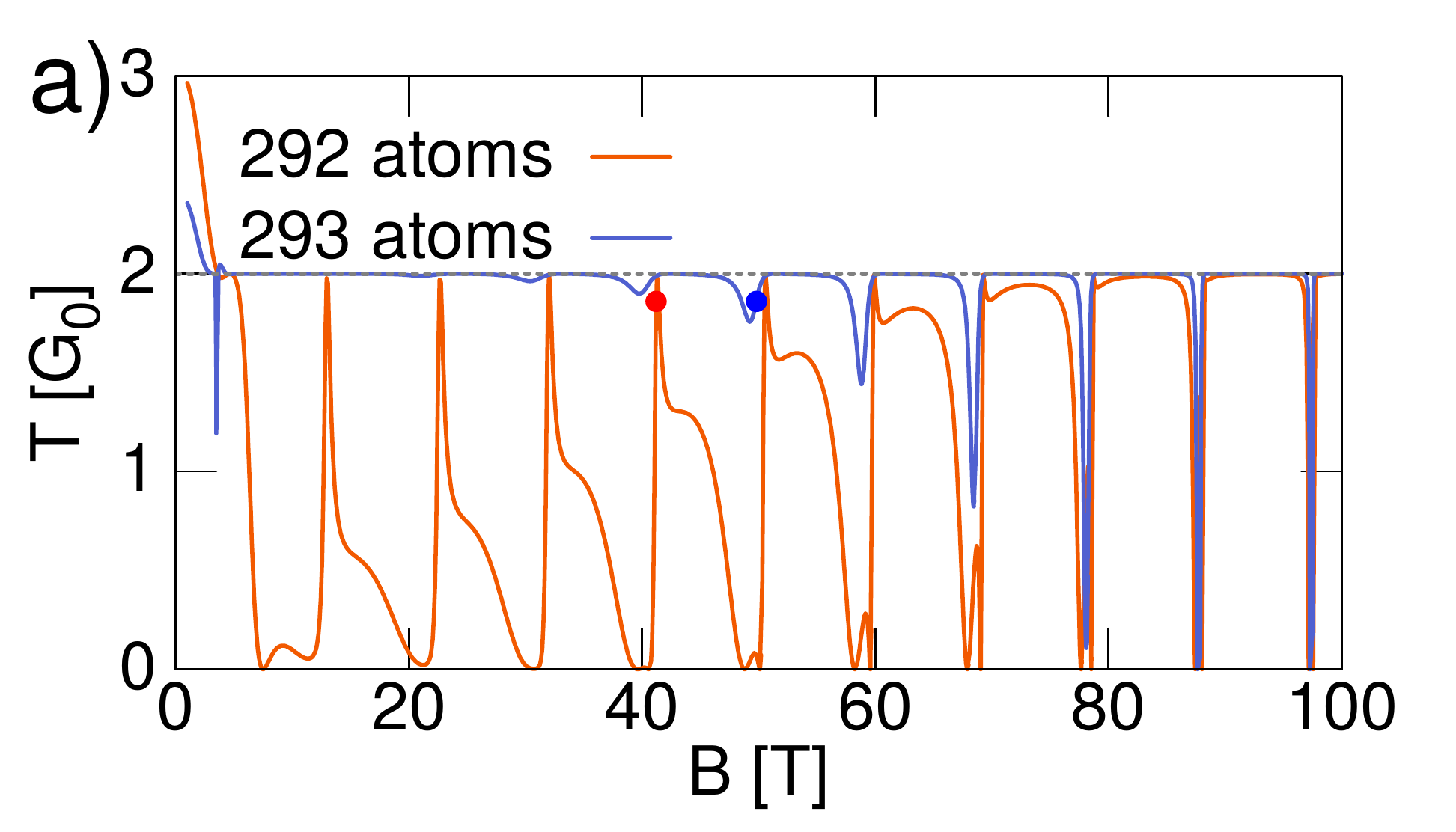}
\includegraphics[width=0.26\paperwidth]{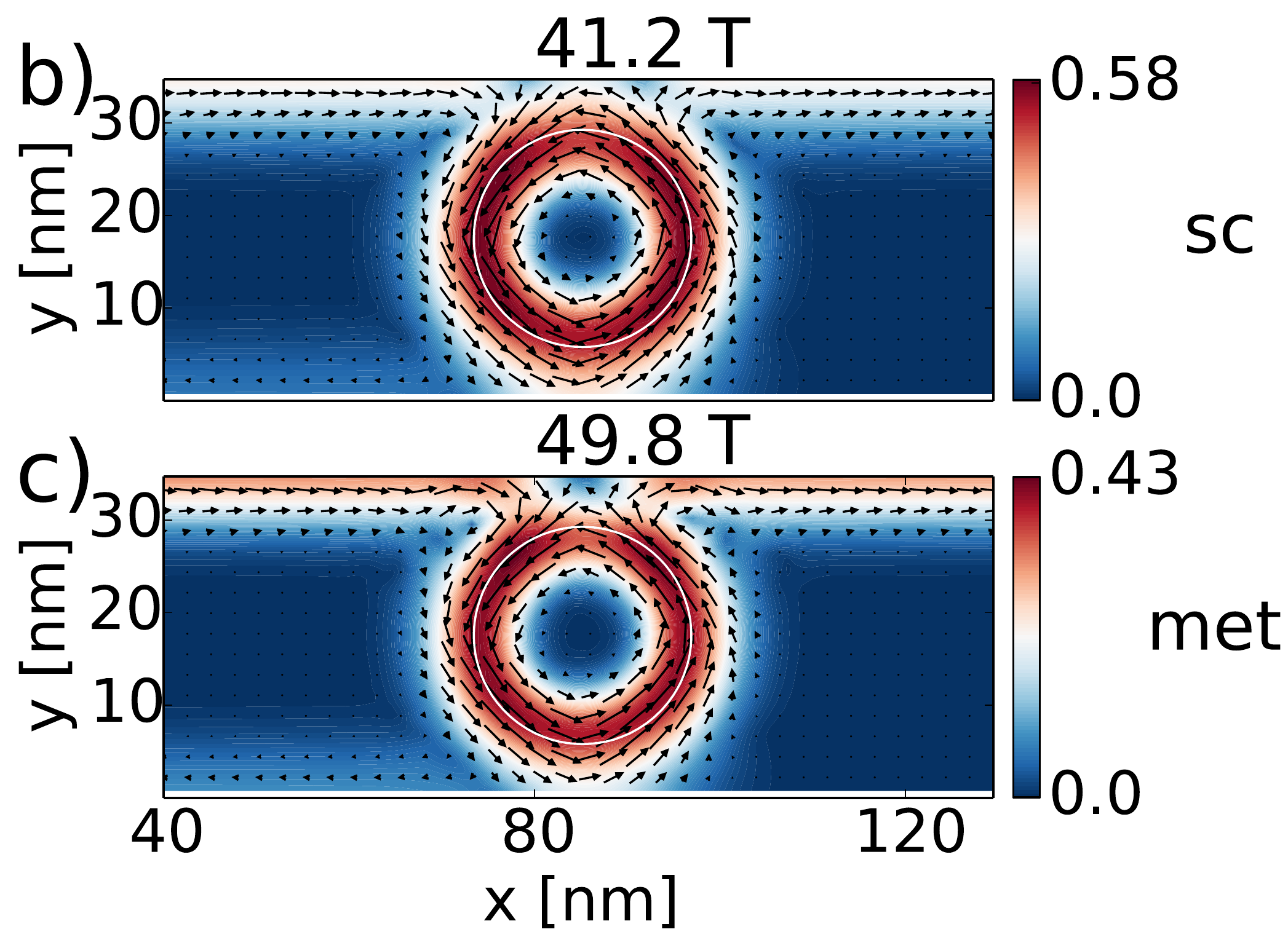}
\caption{
(a) Same as Fig.~\ref{pwB0304}(a) for $E_F=60$ meV. (b,c) square root of the current amplitude [current amplitude calculated using Eq.~(\ref{current})] plotted with orientation of the vector current distribution.
The current plot (b) was made for the semiconducting ribbon -- see the orange dot in (a).
The plot (c) corresponds to the metallic ribbon and was taken for the magnetic field marked by the blue dot in (a).
}\label{pwB0604}
\end{figure*}

Let us first consider the scattering by the tip potential in armchair ribbons in the low energy range when a single subband -- for both current orientations along the ribbon -- is occupied.
We consider nanoribbons of the width of 292 and 293 atoms, which corresponds to 35.79 and 35.92 nm, respectively. At B=0, they are semiconducting and metallic, respectively.
The electron transfer probability as a function of the external magnetic field is displayed in Figs.~\ref{pwB0304}(a) and Fig.~\ref{pwB0604}(a) for the tip
above the center of a semiconducting and a metallic graphene ribbon \cite{Brey} at  $E_F=30$ meV [Fig.~\ref{pwB0304}(a)] and $E_F=60$ meV [Fig.~\ref{pwB0604}(a)] \cite{uwaga}.
The metallic ribbons [blue lines in Fig.~\ref{pwB0304}(a) and Fig.~\ref{pwB0604}(a)] are transparent for low magnetic field. The dependence
on  $B$ -- when it eventually appears above 10 T -- results in a sequence of minima that are periodic in $B$.
For semiconducting ribbons [orange lines in Fig.~\ref{pwB0304}(a) and Fig.~\ref{pwB0604}(a)] the conductance varies with $B$ also in the low-field regime,
but the variation becomes periodic only above 10 T, i.e.~in the quantum Hall regime, where the current starts to flow near the edge and gets confined by the n-p junction.
For higher Fermi energy [$E_F=60$ meV, Fig.~\ref{pwB0604}(a)] at higher $B$ both the semiconducting and
metallic ribbons are nearly transparent for the electron flow outside narrow resonant and periodic dips of conductance.
For low magnetic field in Fig.~\ref{pwB0604}(a) the conductance exceeds $2G_0$ since the filling factor $\nu$ equals 6. As the magnetic field increases, the filling factor drops to 2.

\subsection{Oscillation period}
The oscillation period for a given $E_F$ is similar for both the semiconducting and metallic ribbons [Fig.~\ref{pwB0304}(a) and Fig.~\ref{pwB0604}(a)].
The period dependence on the Fermi energy is distinct  [compare Fig.~\ref{pwB0304}(a) with Fig.~\ref{pwB0604}(a)].
We find that the oscillation period can be quite accurately associated with the radius of the n-p junction induced by the tip,
as given by the $E_F=V(x,y)$ condition [see Fig.~\ref{schemat}(b)]. This condition  produces $R=d\sqrt{V_t/E_F-1}$, which gives $R=17.3$ nm for $E_F=30$ meV and $R=11.7$ nm for $E_F=60$ meV. The AB  oscillation period associated with a ring of radius $R$ is given by $$\Delta B=\frac{h}{e A},$$ with $A=\pi R^2$, and equal to $\Delta B=4.4$ T for $E_F=30$ meV and $\Delta B=9.6$ T for $E_F=60$ meV. This is in good agreement with the periods obtained from the simulation, which are calculated by a Fourier transform of the data (not shown) as $\Delta B=4.7$ T for $E_F=30$ meV and $\Delta B=10$ T for $E_F=60$ meV, for both semiconducting and metallic nanoribbons. The distribution of the current amplitude (given by formula (\ref{current})) is plotted in Figs.~\ref{pwB0304}(b-e) for 7T (b,c) and for the the magnetic fields marked by points (d,e)
in Fig.~\ref{pwB0304}(a). The circles in Figs.~\ref{pwB0304}(b-e) denote the n-p junction line determined by the $E_F=V(x,y)$ condition.
Concluding the above findings, formation of a periodic oscillation pattern that is observed in Fig.~\ref{pwB0304}(a) and  Fig.~\ref{pwB0604}(a) at higher $B$ results from the current confinement at the n-p junction that appears in the quantum Hall regime.

\begin{figure*}[htbp]
\begin{centering}
\includegraphics[width=0.4\paperwidth]{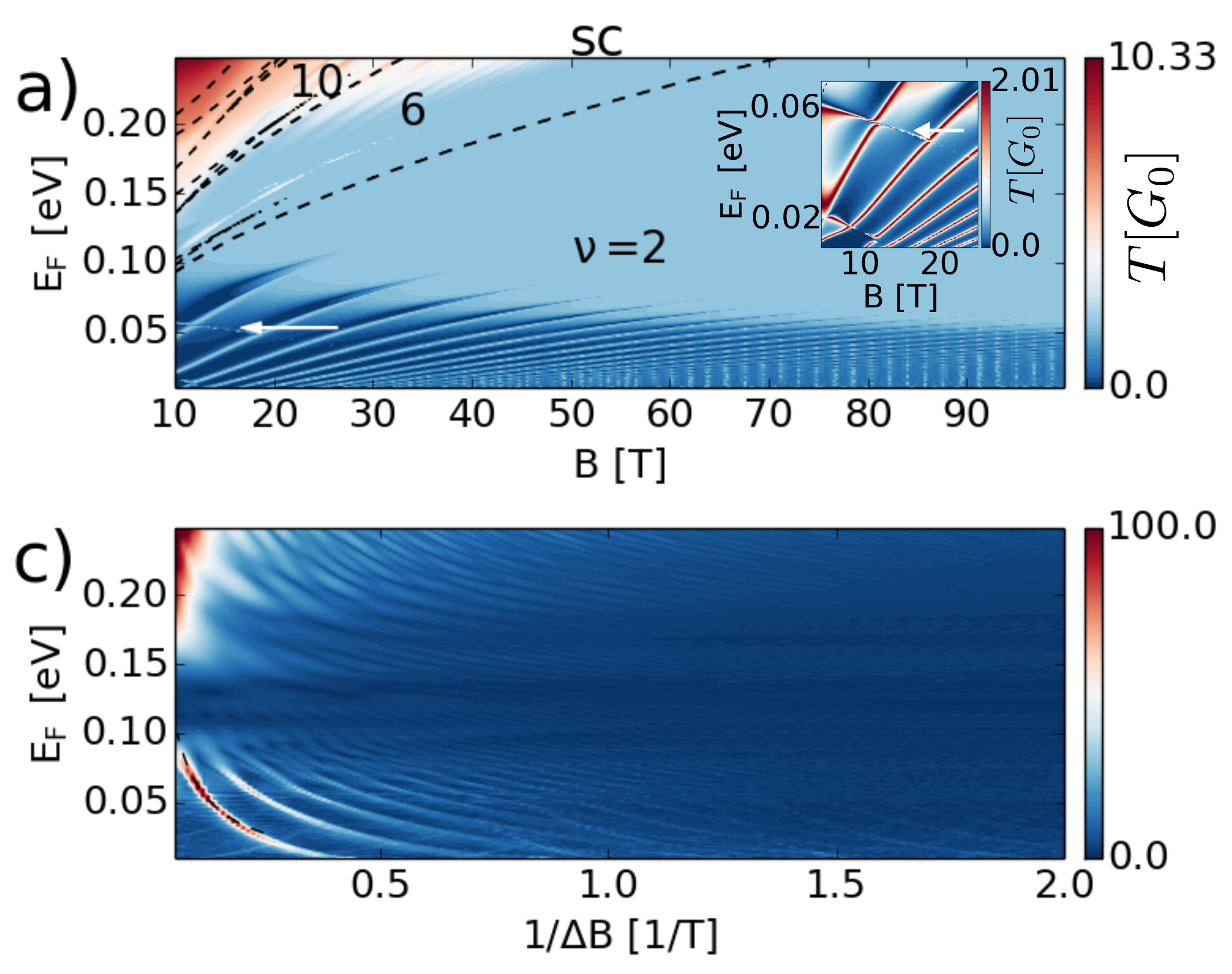}
\includegraphics[width=0.4\paperwidth]{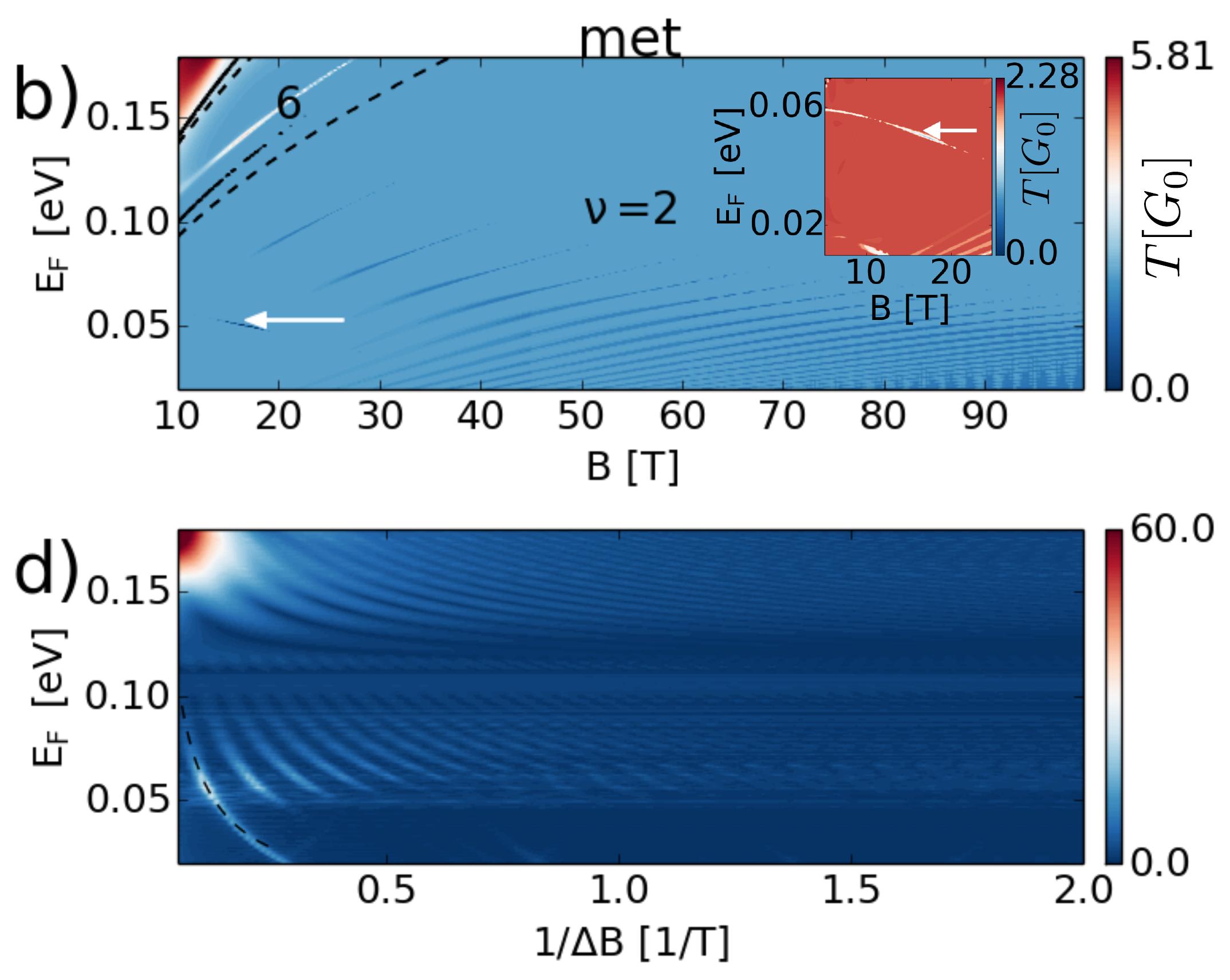}
\includegraphics[width=0.4\paperwidth]{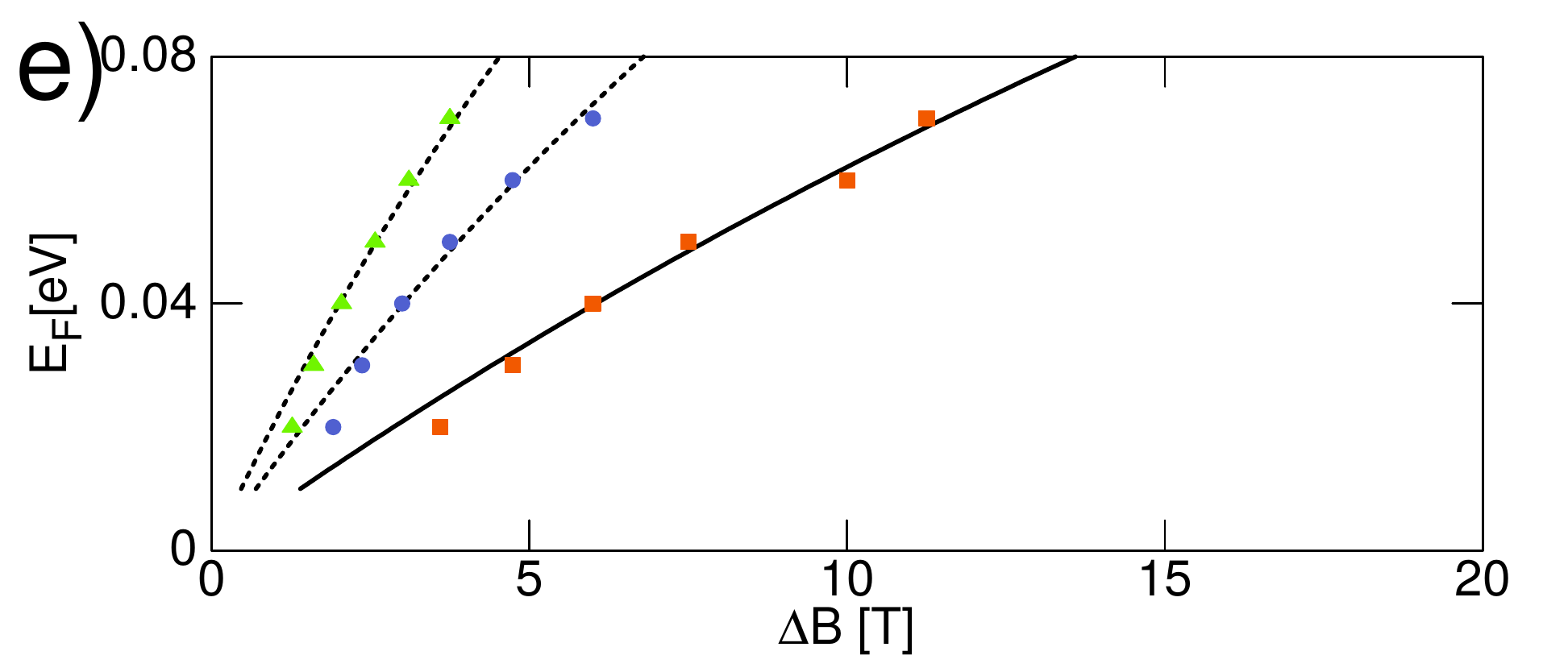}
\includegraphics[width=0.4\paperwidth]{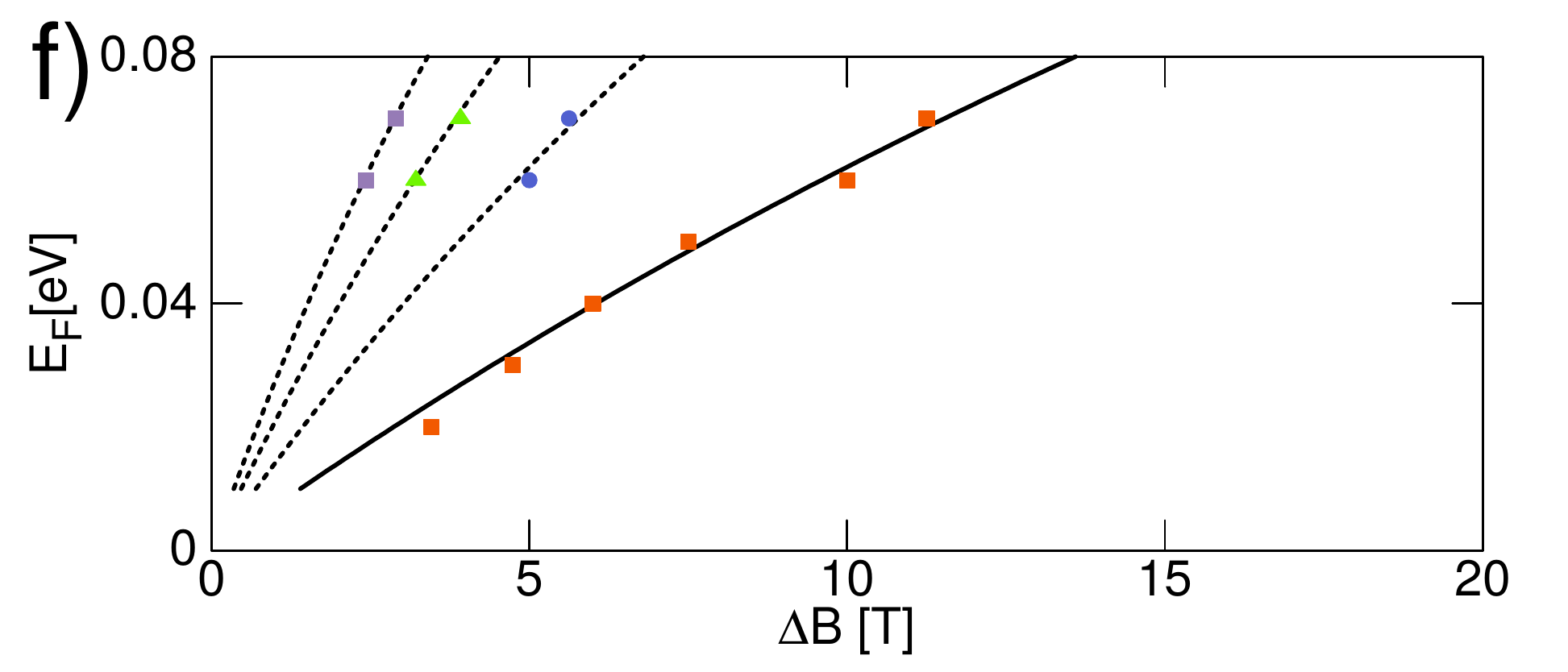}
\par\end{centering}
\caption{\label{wf}
Summed transfer probability for semiconducting (a) and metallic armchair ribbons (b) for parameters as in Fig.~\ref{pwB0304}.
Dashed black lines in (a) and (b) indicate transport threshold for subsequent subbands of the lateral quantization.
The insets show the transfer probability for magnetic field below 25 T.
The arrows in (a) and (b) indicate a feature due to a resonant state localized beneath the tip -- entirely within the p-conductivity region.
The power spectra (Fourier transform)
of the $T(B)$ dependence are displayed in (c) and (d). Dashed black line in (c) and (d) indicate the Aharonov-Bohm period as calculated analytically from the radius of the n-p junction given by the condition $E_F=V(x,y)$.
The numbers in (a,b) denote the filling factor.
In (e,f) the AB period and its 1/2, 1/3, ... fractions calculated for the condition $E_F=V(x,y)$ are shown. The points represent values calculated from several values of the frequencies, at which peaks occur, extracted from the Figs.~(c,d).
}
\end{figure*}

\begin{figure}[htbp]
\includegraphics[width=0.36\paperwidth]{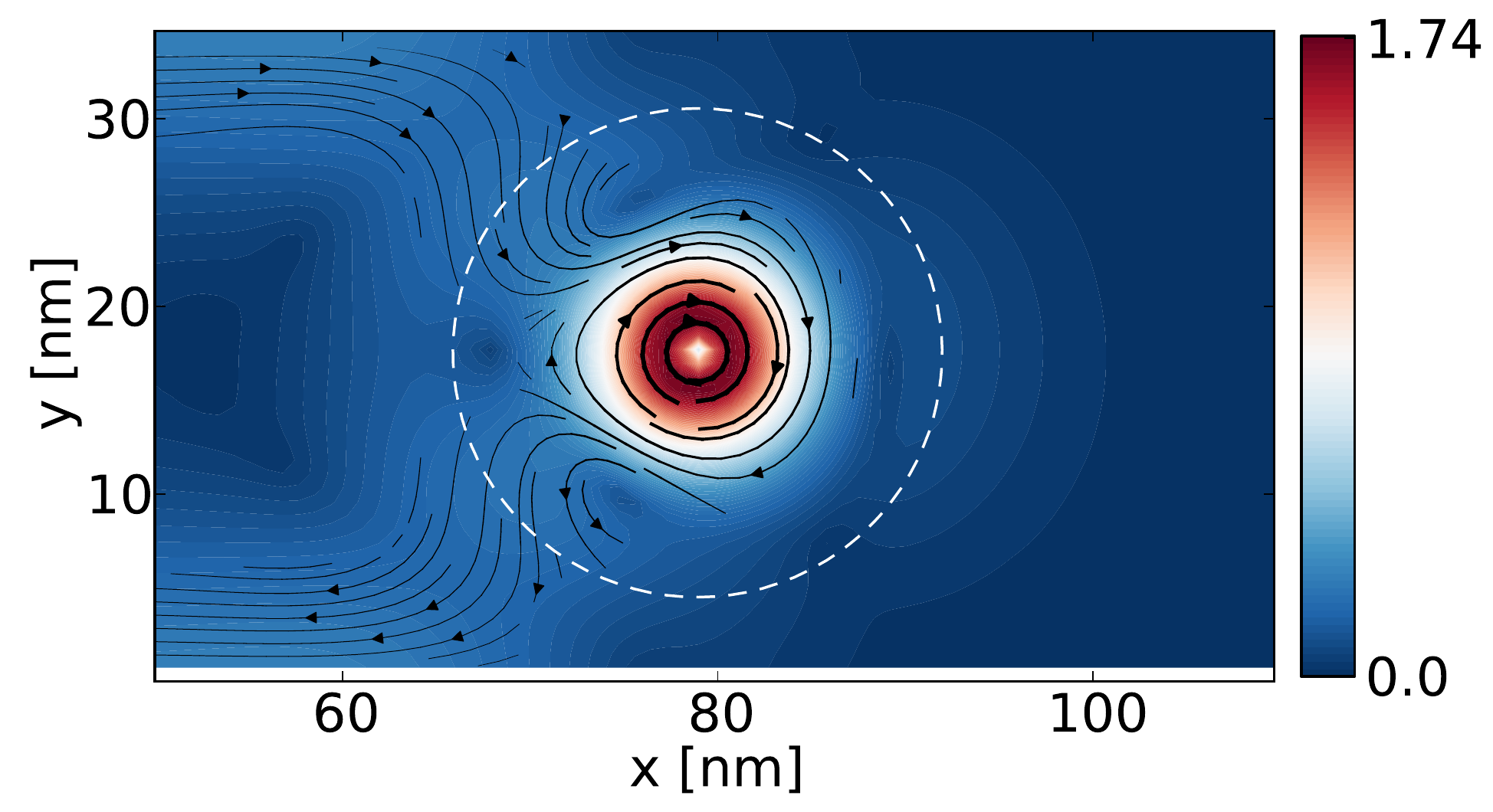}
\caption{The current distribution, obtained using formula (\ref{current}), for the resonance marked by the arrow in Fig.~\ref{wf}(a) for $E_F=50$ meV and $B=16.9$ T.
}\label{reso}
\end{figure}

\subsection{Oscillation amplitude}
%In this section we explain where the difference between metallic and semiconducting armchair nanoribbon come from.
Although the currents in the quantum Hall regime flow near the edges, the conductance in rectangular n-p junctions depends on the width of the nanoribbons. The dependence can be expressed by  the angle between the valley isospins of both edges \cite{Tworzydlo}.
For ribbons with $N$ atoms across, the conductance tends to $2G_0$ (transparent junction)  when $N+1$ is a multiple of 3,
and to $\sfrac{G_0}{2}$ % $^{G_0}/_{2}$
for other $N$ \cite{Tworzydlo}.
At $B=0$  these ribbons happen to be metallic and semiconducting, respectively \cite{Brey}.
According to the present results, in the metallic ribbon the edge current passes smoothly to the circular n-p junction [Figs.~\ref{pwB0304}(c,e)] and goes along
the junction to the other edge of the ribbon. Only a slight reversing current is present at the upper edge of the ribbon near
the n-p junction, in accordance with the aforementioned theory for n-p junctions in metallic ribbons.
In the semiconducting ribbon [Figs.~\ref{pwB0304}(b,d)] the contact of the n-p junction and the edge of the ribbon acts as a beam
splitter \cite{bms}.
The contact at the lower edge of the ribbon backscatters part of the current to the input lead.
The beam splitter at the upper contact right of the tip sends a part of the current to the right output lead, and keeps another part circulating
around the junction. In consequence, we have a pronounced current going all around the ring [the upper edge-junction contact in Fig.~\ref{pwB0304}(d) and Fig.~\ref{pwB0604}(b)].
As seen in Fig.~\ref{pwB0304}(a), the visibility of the conductance oscillations for
the semiconducting ribbon is much higher than those for the metallic ribbon, with only a residual current at the upper
junction-edge contact.

For the metallic ribbon, the amplitude of the conductance oscillations grows with increasing magnetic field
%for higher Fermi energy
 [the blue line in Fig.~\ref{pwB0604}(a)],
which is accompanied by a growth of the oscillation period, and a reduction of the width of the conductance minima with increasing Fermi energy.
For increasing Fermi energy, the radius of the junction is reduced [compare Fig.~\ref{pwB0604}(c) and Figs.~\ref{pwB0304}(c,e)],
and thus the coupling of the edge current to the junction current gets weaker. The circular currents confined at the junction
form resonant states with long lifetimes, hence the abrupt form of the conductance oscillation at high magnetic field [cf. the dips at Fig.~\ref{pwB0604}(a)],
when the shift of the scattering density to the edge of the ribbon -- as due to the classical Lorentz force  \cite{LF} -- is stronger.
Outside these resonances the presence of the tip potential does not influence the electron transfer probability [Fig.~\ref{pwB0604}(a)].
 In the experimental conditions one can manipulate alternatively the Fermi energy (by a back gate voltage \cite{xliu}) or the potential applied to the tip.

\subsection{Edge-junction coupling and power spectra}
In Fig.~\ref{wf} the summed transfer probability of current from source to drain is plotted as a function of both $E_F$ and $B$
for the semiconducting [Fig.~\ref{wf}(a)] and metallic [Fig.~\ref{wf}(b)] ribbons.
%, with the corresponding power spectra displayed in Fig. \ref{wf}(c) and Fig. \ref{wf}(d).
The power spectra in Fig.~\ref{wf}(c) (Fig.~\ref{wf}(d)) show the Fourier transform of each cross-section $T(B)$ of
Fig.~\ref{wf}(a) (Fig.~\ref{wf}(b)) as a function of $E_F$.
For the semiconducting ribbon the conductance oscillations disappear above
the lowest-subband region (Fig.~\ref{wf}(a), $E_F>0.1$ eV), and for the metallic ribbon [Fig.~\ref{wf}(b)] formation
of the AB conductance oscillations is shifted towards higher $B$.

The edge current that is coupled to the n-p junction by the tip forms a system geometrically related to a quantum dot \cite{zitko} or a quantum ring \cite{minchullee} singly connected
to the channel. The conductance of these systems is governed by Fano interference effects, which for quantum rings \cite{poniedzialek} produce
resonances of width (lifetime) which is reduced or increased by the external magnetic field depending on the orientation of the current circulation around the ring.
The stabilization of the resonant lifetime \cite{poniedzialek} is due to the classical Lorentz force which modifies localization of the currents at the edges of the sample.
The orientation of the circulating current determines the projection
of the produced magnetic moment with the external  magnetic field leading to a growth or reduction of the
resonance energy with B [60], for the produced magnetic dipole aligned antiparallel or parallel to the
external magnetic field vector, respectively.
In the present results all periodic structures in conductance become thinner at high magnetic field, and the energies of the lines grow with the magnetic field
since the dipole moment generated by the current in the resonant states is opposite to the external magnetic field. In contrast to the rings with tailored confinement \cite{poniedzialek} no periodic lines with energies that fall in $B$
  and increase in width are found.  This is because the n-p junction at $B>0$ supports confinement of counterclockwise currents only, while
  the tailored rings host currents of both orientations.

Similar current distributions were obtained in Ref.~\onlinecite{Rakyta}, where the circular hole in the ribbon at high magnetic field supports current circulating in the direction forced by the magnetic field.
At higher magnetic fields [Fig.~3 of Ref.~\onlinecite{Rakyta}] the resonant energy levels
start to increase linearly with the external magnetic field, and the energy spacings between the levels become equidistant. Hence, the resonant energy levels appear at a fixed Fermi energy periodically, and the period seen in Fig.~3 of Ref.~\onlinecite{Rakyta} is equal or close to the flux quantum.

In Figs.~\ref{wf}(a,b) we spot a single distinct line that falls with $B$: with the energy of $E_F=50$ meV near $B=15$ T (see the arrows in Figs.~\ref{wf}(a,b)]. This line corresponds
to a resonance trapped beneath the tip, in the $p$-conductivity region. The current in this resonance circulates clockwise, and the Lorentz force
tends to localize it strictly beneath the tip (see Fig.~\ref{reso}).
The position of the resonance and the current circulation is the same both for metallic and semiconducting ribbons [cf. Figs.~\ref{wf}(a) and (b)].
Note, that recently -- similar resonant states localized under the scanning tunneling microscope tip within a circular n-p junction induced by the tip voltage -- but in the absence of external magnetic field,
where detected by the tunneling currents between the graphene and the microscope probe.

The resonant states trapped in the $p$-region beneath the tip [lines marked by arrows in Figs.~\ref{wf}(a,b)] are related to the tip and are independent of the type of the ribbon edge,
although the resonance is more pronounced for the metallic ribbon [Fig.~\ref{wf}(b)], for which no AB oscillations are present in the region where the resonance is observed.
At higher field the resonance line disappears from the conductance spectra of both metallic and semiconducting ribbons.
The disappearance of the line is not due to dissolution of the resonance,
but to the fact that the resonant states gets isolated from the bulk of the sample by the external counterclockwise current loop along the n-p junction.

The Fourier transform of the $T(B)$ dependence calculated for the armchair ribbon [Fig.~\ref{wf}(c)] below $E_F<0.1$ eV indicates the variation
of the Aharonov-Bohm period with the Fermi energy. %, which very well agrees with the condition $E_F=V(x,y)$ discussed above.
The black dashed line at the left hand side of the plot marks the elementary Aharonov-Bohm period, as calculated analytically. The power spectrum contains also higher harmonics (integer multiples of the fundamental frequency).
In Figs.~\ref{wf}(e,f), we give for several $E_F$ the values of $\Delta B$ calculated from the frequencies at which the peaks occur, extracted from the Figs.~\ref{wf}(c,d), together with the analytical values of the fundamental period (black solid line) and its $1/2$, $1/3$, ... fractions representing higher harmonics (black dashed lines) calculated for $R$ given by the condition $E_F=V(x,y)$. In case of a semiconducting ribbon [Fig.~\ref{wf}(e)] we have extracted up to 3 values since the following ones were difficult to distinguish from the background. Up to the energy 28 meV, the diameter of the circle $E_F=V(x,y)$ is larger than the ribbon width, and the area encircled by the current is smaller than the area of this circle, hence the deviation from the analytical values. However, qualitatively  the expected period is bigger than the one of a circular current path, which matches our observations. Above the energy 28 meV, the entire circle is inside the ribbon, and the analytical and modelled values are in a good agreement. Similar observations can be made for fundamental frequencies extracted for the metallic ribbon [Fig.~\ref{wf}(f)].
The higher harmonics appear above the energy 60 meV, thus the entire junction is located inside the ribbon and the extracted points lie nearly perfectly on the analytical lines.

For the metallic ribbon the higher harmonics appear only for $E_F>50$ meV, and for lower Fermi energy only the elementary period is observed.
The higher harmonics are present when $T(B)$ minima turn into abrupt dips [Fig.~\ref{pwB0604}(a)] which relates to formation of
resonances with multiple loops performed by the flowing electron around the n-p junction \cite{Spivak, colinPRB}. For stronger coupling of the junction to the edge, which appears
at lower Fermi energy, the transfer probability is determined by the interference between  the current that circulates below the junction and the residual one which goes straight at the upper
edge [Fig.~\ref{pwB0304}(e)]. This interference corresponds to the one-pass conditions discussed in the original paper of Aharonov and Bohm \cite{AB}, and produces the sine-form dependence of $T(B)$.
Only for higher Fermi energy the current starts to circulate around the ring-like n-p junction [Fig.~\ref{pwB0604}(c)],
with the phase accumulated from the vector potential proportional to the number of turns, and higher harmonics appear in the power spectrum [Fig.~\ref{wf}(d)].
For the semiconducting ribbon, the higher harmonics are present also at low energy [Fig.~\ref{wf}(b)] due to the beam-splitting role of the junction/edge contacts [Fig.~\ref{pwB0304}(d)].

We studied the stability of the results against intervalley scattering, due to e.g.~missing atoms in the ribbon or point potential defects \cite{defects}.
 For a moderate number of defects, the effects for a semiconducting armchair ribbon is weak. This can be understood due to the fact that
the intervalley scattering is inherently present in semiconducting armchair ribbons \cite{WurmSym}.
The transmission amplitude is more strongly affected in the metallic armchair ribbon. 
For the metallic armchair ribbons the AB oscillations survive.
The effects due to disorder are more pronounced at low magnetic field. 

\begin{figure}[htbp]
\begin{centering}
\includegraphics[width=0.2\paperwidth]{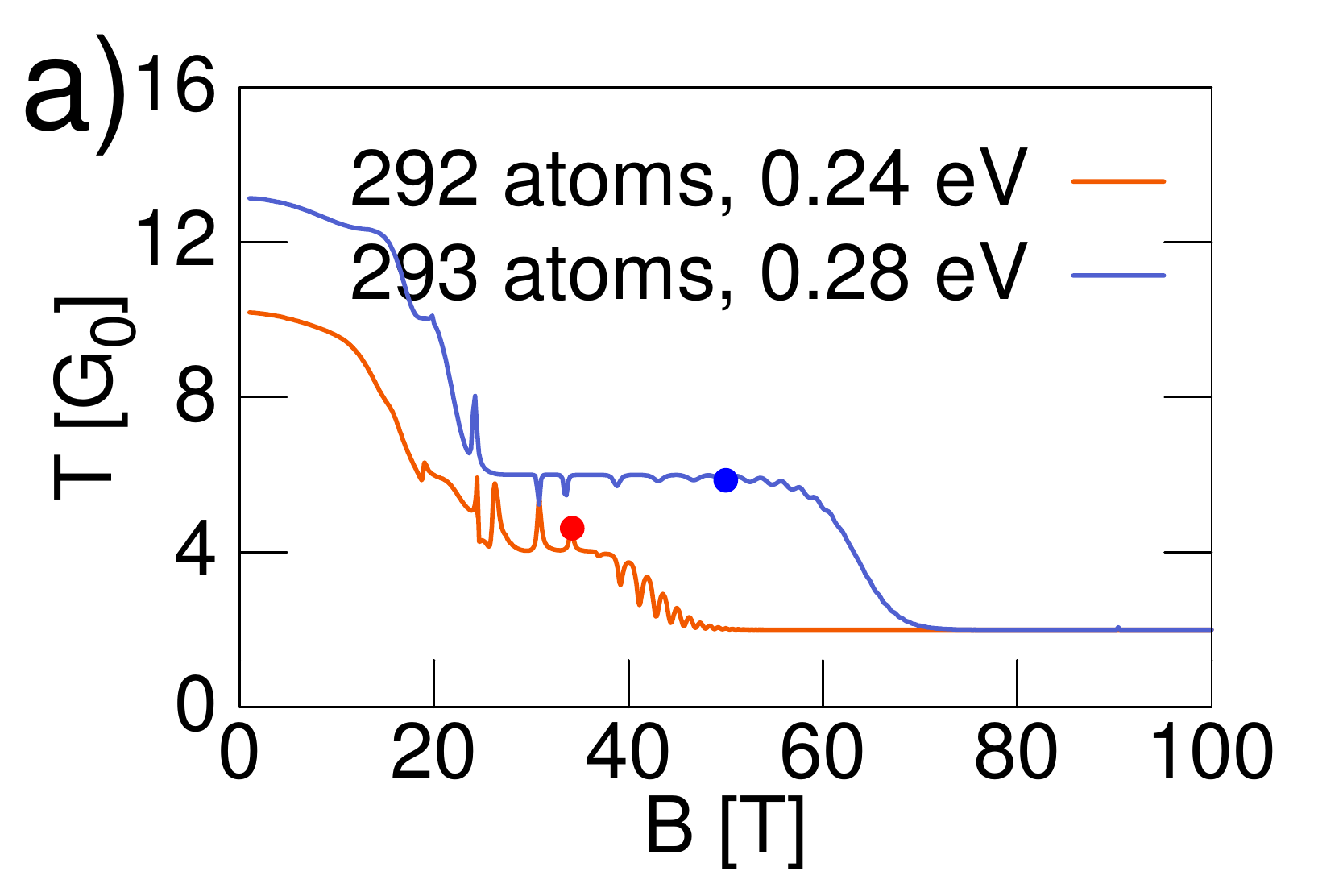}
\includegraphics[width=0.16\paperwidth]{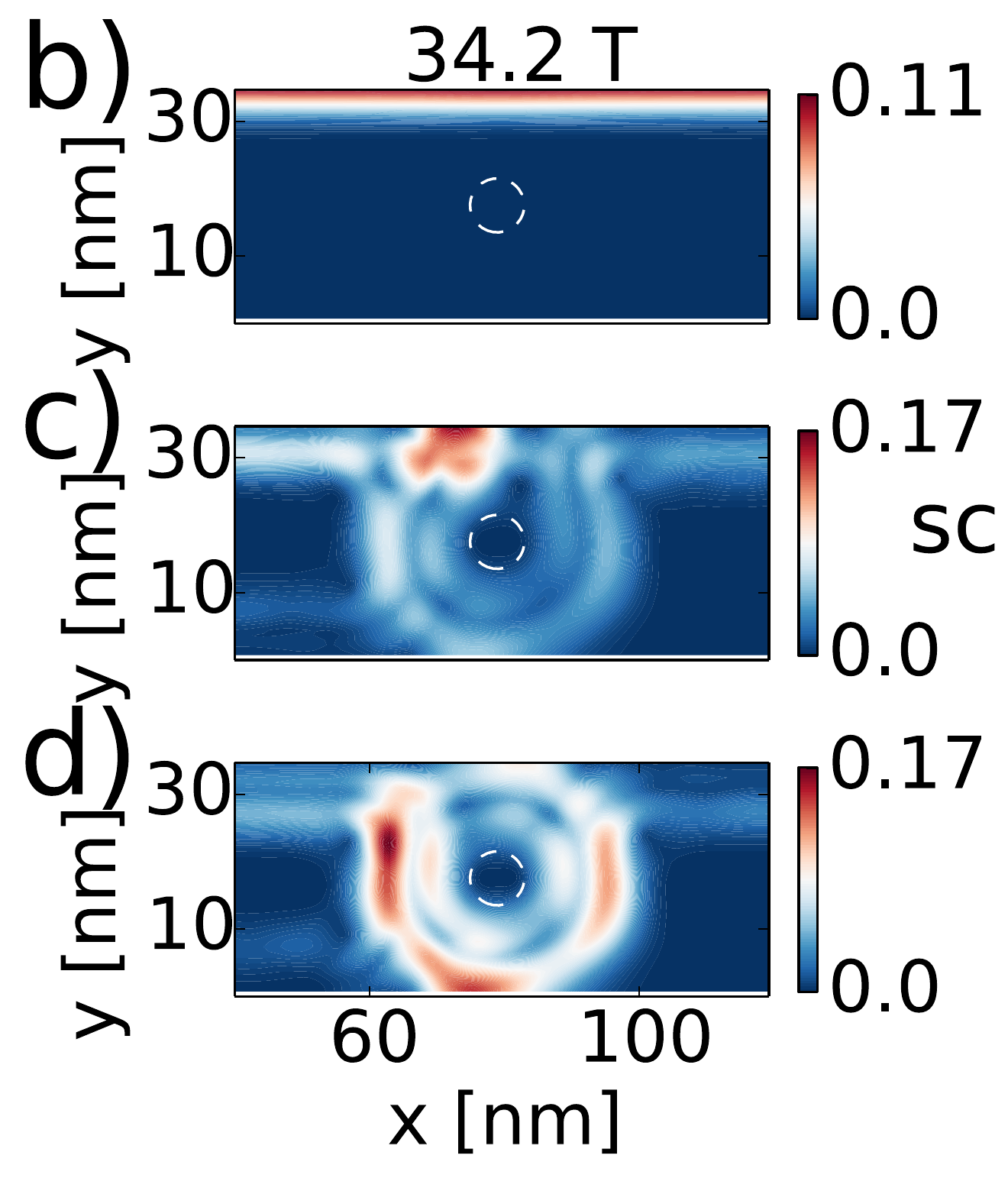}
\includegraphics[width=0.4\paperwidth]{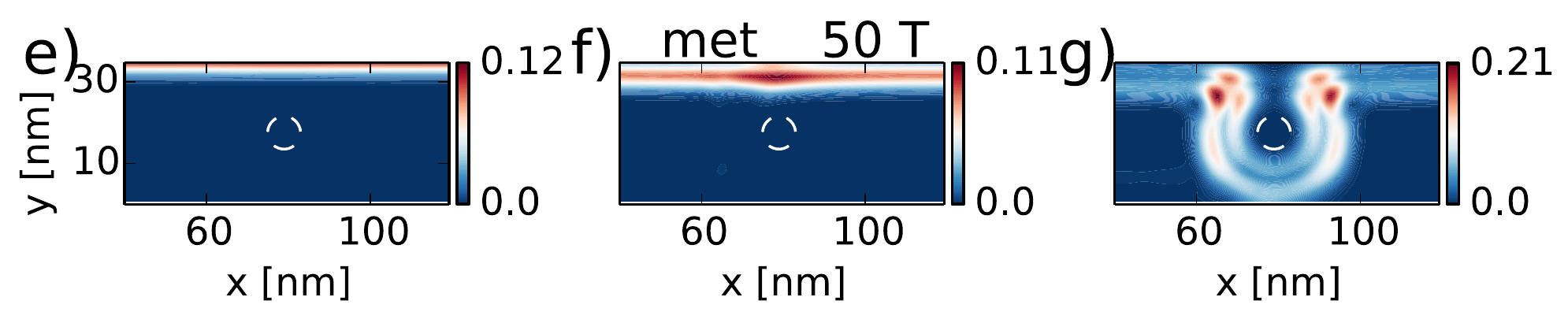}
\par\end{centering}
\caption{\label{wye}
(a) Transfer probability versus magnetic field for the semiconducting (orange line) and metallic (blue line) ribbons of Figs.~\ref{wf}(a,b) -- cross sections taken at $E_F=240$ meV for
the semiconducting and $E_F=280$ meV for the metallic ribbon.
(b-g) Current amplitude calculated using Eq.~(\ref{current}) with a wave function of a single subband.
(b-d) Current amplitude for the lowest, second, and third subband for the semiconducting ribbon at  $B=34.2$ T [the orange dot in (a)].
(e-g) Current amplitude for the lowest, second, and third subband for the metallic ribbon at  $B=50$ T [the blue dot in (b)].
The circles in (b-g) indicate the nominal position of the n-p junction defined as the place in space where the potential becomes
equal to the Fermi energy.
}
\end{figure}

\subsection{Multiple conducting subbands case}
Let us consider the conductance at the higher energy -- for filling factor higher than 2. The summed transfer dependence is plotted in Fig.~\ref{wye}(a).
For fixed Fermi energy the conductance drops with $B$, since the number of conducting subbands is reduced. Oscillations of conductance in the magnetic
field are observed for three subbands at the Fermi level. The oscillations are not perfectly periodic in $B$.  Figures \ref{wye}(b-d) present
the probability current densities for the electron incident from the lowest, second, and third Landau levels, respectively. Electrons from the lowest Landau level ignore the presence of the tip,
and current circulation is present only for the other two Landau levels. The current loops are doubled: two circular features appear one beside the other. In the second Landau level [Fig.~\ref{wye}(c)]
a resonance is observed near the upper edge-junction contact. In the plots of Figs.~\ref{wye}(b-d)
the current flows entirely in the n-region.
For the metallic ribbon, the first and second subbands pass above the tip, Figs.~\ref{wye}(e-f), and the third bypasses it below, Fig.~\ref{wye}(g), with only a weak current near
the upper edge above the tip.
Summarizing, for the armchair ribbons, clear and ideally periodic AB oscillations are only observed in the lowest subband transport conditions: for filling factor $\nu=2$ at the n-conductivity region.
The graphene structures exhibit a sufficient tunability of the Fermi energy, local potential, and the filling factors \cite{ob} to set the workpoint for the AB conductance oscillations.

\section{Summary and Conclusion}
We have studied the transport properties of graphene armchair nanoribbons with a closed n-p junction induced electrostatically by a gate floating above the sample
using the quantum transmitting boundary method for the tight-binding Hamiltonian.
The system acquires conductance oscillations of the Aharonov-Bohm periodicity in the quantum Hall regime
with a Fermi-energy-dependent period.
The conductance periodicity is due to the current confinement along the n-p junction induced in high magnetic field and the coupling
of the junction currents to the current flowing along the edges of the ribbon.
Observation of the Aharonov-Bohm interference requires a beam-splitting behaviour of the
junction/edge contacts which we find for the ribbons with armchair edges at low Fermi energy.
For the metallic armchair ribbons at low Fermi energy the power spectrum of conductance possesses the lowest harmonics only.
The discussed circular n-p junction supports only confinement of currents producing
magnetic dipole moments opposite to the external magnetic field.
We find that the resonant states localized under the tip
with opposite current circulation get isolated from the bulk of the sample at high magnetic field by a loop of n-p currents
flowing along the n-p junction.

\section*{Acknowledgments}
This work was supported by National Science Centre (NCN)
according to decision DEC-2015/17/B/ST3/01161, and by
PL-Grid Infrastructure. The first author is supported by the
scholarship of Krakow Smoluchowski Scientific Consortium
from the funding for National Leading Reserch Centre by
Ministry of Science and Higher Education (Poland)
as well as by Natonal Science Centre (NCN) Etiuda scholarship
DEC-2015/16/T/ST3/00264.

We acknowledge funding by the executive programme for scientific and technological cooperation between the Italian Republic and the Republic of Poland for the years 2016 - 2018.

SH acknowledges funding from the European Union Seventh Framework Programme under Grant Agreement No.~604391 Graphene Flagship. Furthermore, financial support from the CNR in the framework of the agreements on scientific collaboration between CNR and JSPS (Japan), CNRS (France), and RFBR (Russia) is acknowledged. He also acknowledges funding from the Italian Ministry of Foreign Affairs, Direzione Generale per la Promozione del Sistema Paese.


\begin{thebibliography}{10}
\bibitem{defects} The defects are introduced as vacancies, with a random removal of carbon atoms in a region of width 1 nm and length 55 nm , located near the np junction, close to the upper ribbon edge (probability of atom removal 0.02).
\end{thebibliography}
\end{document}